\documentclass[journal,12pt,draftclsnofoot,onecolumn,english]{IEEEtran}

\usepackage{amssymb}
\usepackage{amscd}
\usepackage{amsmath}
\usepackage{epsfig}
\usepackage{graphics}
\usepackage{psfrag}
\usepackage{rotating}
\usepackage{amsmath}
\usepackage{amsfonts}
\usepackage{url}
\usepackage{color}
\usepackage{subcaption}

\usepackage{epstopdf}
\usepackage{amsthm}
\usepackage{tikz}
\usepackage{mathtools}
\usepackage{multirow}
\usetikzlibrary{arrows,shapes,decorations,backgrounds}
\usepackage[final]{pdfpages}
\usepackage{enumitem}
\usepackage[binary-units]{siunitx}
\DeclareSIUnit{\belmilliwatt}{Bm}
\DeclareSIUnit{\belsquaremeter}{Bsm}
\DeclareSIUnit{\dBm}{\deci\belmilliwatt}
\DeclareSIUnit{\dBsm}{\deci\belsquaremeter}
\usepackage{cite}

\newcommand{\Na}{N_{\rm a}}
\newcommand{\Nrf}{N_{\rm rf}}

\usepackage{multicol}

\long\def\comment#1{}

\newfont{\bbb}{msbm10 scaled 700}

\newfont{\bb}{msbm10 scaled 1100}
\newcommand{\CC}{\mbox{\bb C}}

\newcommand{\EE}{\mbox{\bb E}}

\newcommand{\av}{{\bf a}}

\newcommand{\fv}{{\bf f}}

\newcommand{\sv}{{\bf s}}

\newcommand{\wv}{{\bf w}}
\newcommand{\vv}{{\bf v}}
\newcommand{\xv}{{\bf x}}
\newcommand{\yv}{{\bf y}}

\newcommand{\Bm}{{\bf B}}

\newcommand{\Dm}{{\bf D}}

\newcommand{\Fm}{{\bf F}}

\newcommand{\Hm}{{\bf H}}
\newcommand{\Id}{{\bf I}}

\newcommand{\Rm}{{\bf R}}

\newcommand{\Um}{{\bf U}}

\newcommand{\Oc}{{\cal O}}

\newcommand{\phiv}{\hbox{\boldmath$\phi$}}
\newcommand{\psiv}{\hbox{\boldmath$\psi$}}

\newcommand{\Gammam}{\hbox{\boldmath$\Gamma$}}

\newcommand{\Psim}{\hbox{\boldmath$\Psi$}}

\renewcommand{\arg}{{\hbox{arg}}}

\newcommand{\SNR}{{\sf SNR}}

\newcommand{\eqdef}{\stackrel{\Delta}{=}}

\newcommand{\T}{{\scriptscriptstyle\mathsf{T}}}
\renewcommand{\H}{{\scriptscriptstyle\mathsf{H}}}

\begin{document}

\title{Beam Refinement and User State Acquisition via Integrated Sensing and Communication with OFDM}
\author{Fernando~Pedraza, Mari~Kobayashi~and~Giuseppe~Caire
\thanks{
	F.~Pedraza and G.~Caire are with the Electrical Engineering and Computer Science Department, Technische Universit\"at Berlin, 10587 Berlin, Germany (email: f.pedrazanieto@tu-berlin.de; caire@tu-berlin.de )
}
\thanks{
	M.~Kobayashi is with the Electrical and Computer Engineering Department, Technische Universit\"at M\"unchen, 80333 Munich, Germany (email: mari.kobayashi@tum.de)
}
}
\maketitle

\begin{abstract}

The performance of millimeter wave (mmWave) communications strongly relies on accurate beamforming both at base station and user terminal sides, referred to as beam alignment (BA). Existing BA algorithms provide initial yet coarse angle estimates as they typically use a codebook of a finite number of discreteized beams (angles). Towards emerging applications requiring timely and precise tracking of dynamically changing state of users, we consider a system where a base station with a co-located radar receiver estimates relevant state parameters of users and simultaneously sends OFDM-modulated data symbols. In particular, based on a hybrid digital analog data transmitter/radar receiver architecture, we propose a simple beam refinement and initial state acquisition scheme that can be used for beam and user location tracking in a dynamic environment. 
Numerical results inspired by IEEE802.11ad parameters demonstrate that the proposed method is able to improve significantly the communication rate and further achieve accurate state estimation. 

\end{abstract}


\section{Introduction}
The integrated sensing and communication at millimeter wave (mmWave) or higher frequency bands has been considered as one of key enablers for future cellular communication standards. The efficient operation at such high frequency bands strongly relies on accuracy and timely beamforming at a base station (BS) and user terminals (UEs), referred to as {\it beam alignment}, in order to compensate high propagation and penetration losses (see e.g. \cite{Heath-SparsePrecoding}). 
The beam alignement (BA) design achieving a good tradeoff between alignment accuracy and required resource overhead has been extensively studied in the literature (e.g. \cite{Xiaoshen} and references therein). The existing BA algorithms establish the best beam pair for each BS-UE link from a codebook consisting of a finite number of discretized beams after a suitable synchronization procedure \cite{Xiaoshen, Heath-beamforming}. Unfortunately, the performance of such a codebook-based approach is limited by inherent angle discretization errors, which yields non-negligible errors for emerging applications such as vehicular to everything (V2X) requiring precise localization. Moreover, these applications typically build on timely and precise tracking of dynamically changing {\it state} of UEs, determined by the angle, delay, and Doppler, as studied as beam tracking problem in \cite{BeamTracking1, JRC-survey}. These observations motivate us to study the beam refinement and state acquisition by assuming that initial beam acquisition and synchronization between the BS and the UEs are already performed. Notice that our approach opportunistically exploits existing communication protocols to enhance the communication and sensing performances. Hence, it is conceptually different from sensing-aided BA that aims to speed up the initial beam acquisition by exploiting radar or other side information (see e.g. \cite{gonzalez2016radar2,va2017position2}). 

In this paper, we address a downlink scenario where a BS, equipped with a co-located radar receiver and multiple antennas/RF chains, wishes to send multiple data streams to UEs using orthogonal frequency division multiplexing (OFDM) and spatial beamforming. In order to incorporate coarse angle estimates available at the BS, we consider a simple hybrid digital analog (HDA) architecture consisting of a set of tunable phase shifters followed by a fixed bank of spatial filters (or a reduction matrix) to output the observation corresponding to the number of RF chains. Under this setup, we propose a beam refinement and state acquisition scheme. By focusing on a semi-unitary structure of the reduction matrix, our proposed scheme adapts the phase shifters to the initial angle estimates and obtains refined angle, delay, and Doppler shift estimation. Numerical results inspired by IEEE802.11ad parameters demonstrate that the proposed method is able to improve significantly the communication rate and further achieve accurate state estimate for various levels of angle discretization errors. 

\section{System Model}

\begin{figure}[!t]
	\centering
	\includegraphics[width=0.45\textwidth]{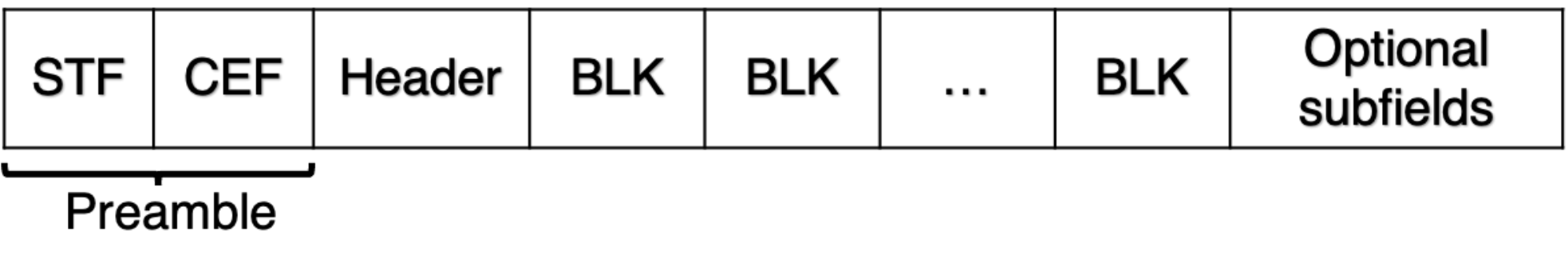}
	\caption{IEEE802.11ad frame structure consisting of a short training field, channel estimation field, headers and data communication blocks.}
	\label{fig:Frame}
\end{figure}

\begin{figure}[!t]
	\centering
	\begin{subfigure}[b]{.5\textwidth}
		\centering
		\includegraphics[width=\textwidth, height=.2\textheight]{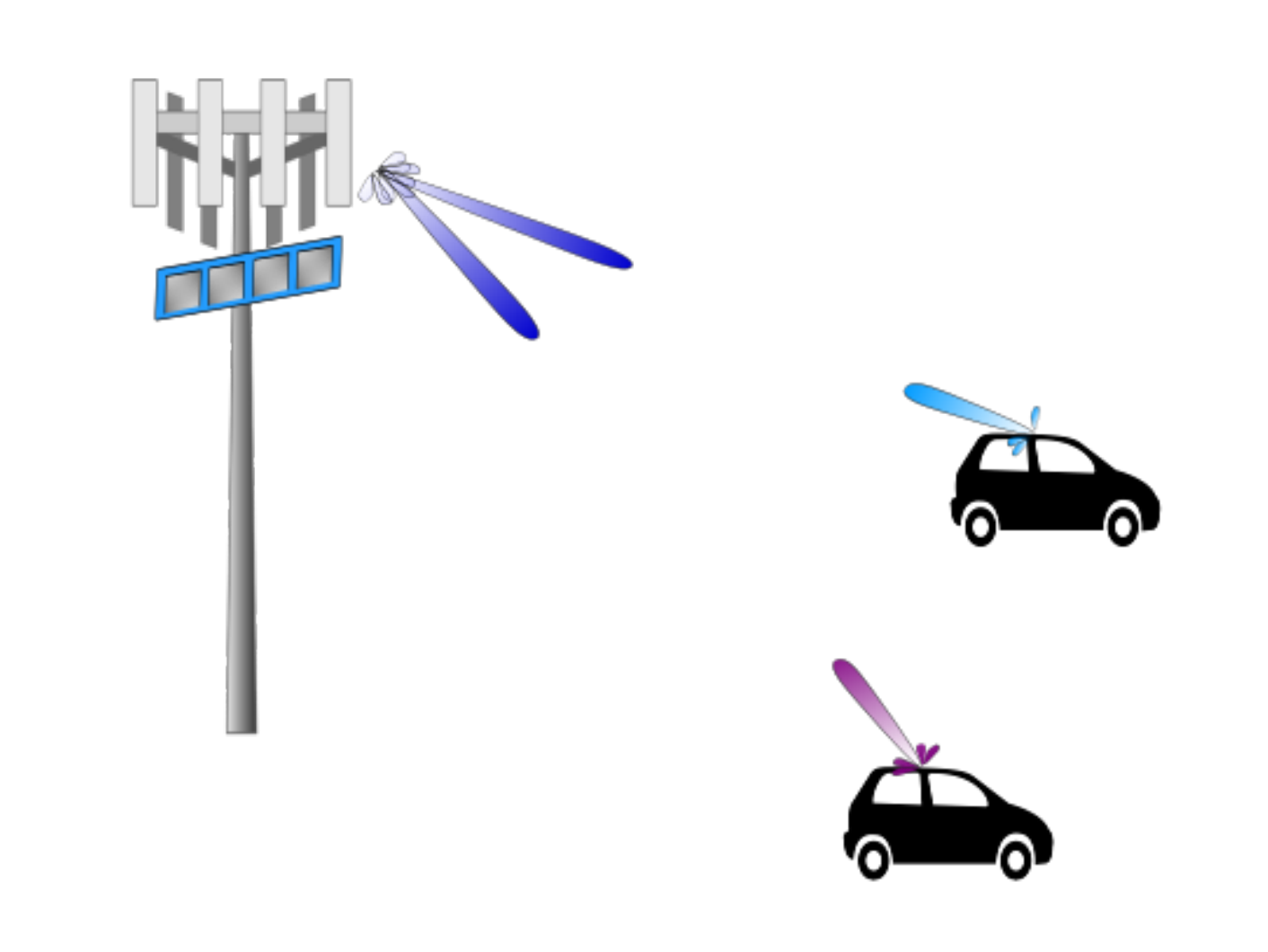}
		\caption{BA output}
		\label{fig:SuccessfulBA}
	\end{subfigure}
	\begin{subfigure}[b]{0.4\textwidth}
		\centering
		\includegraphics[width=\textwidth, height=.2\textheight]{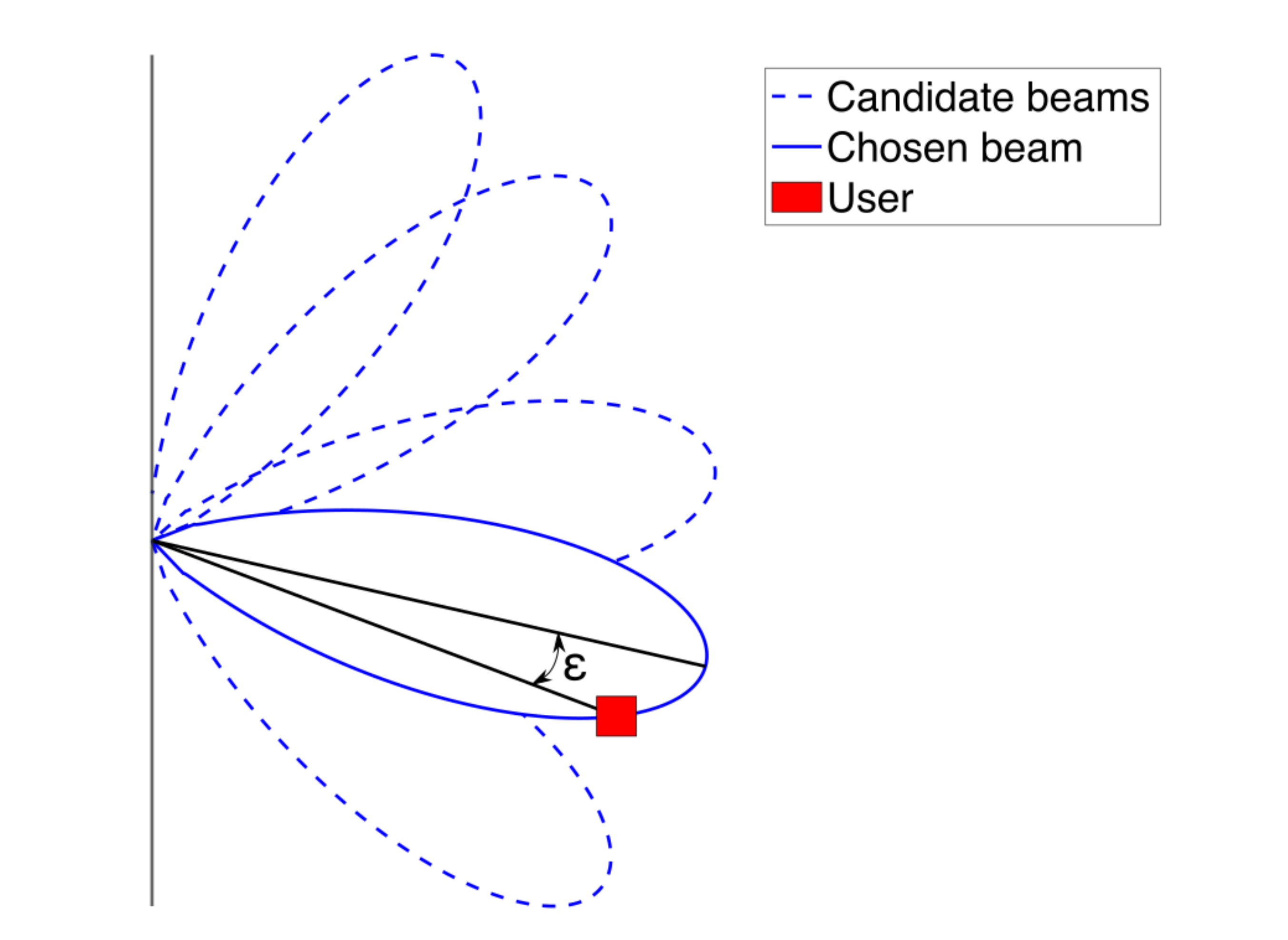}
		\caption{Angle discretization error $\epsilon$}
		\label{fig:BeamRefinementExample}
	\end{subfigure}
	\caption{Illustration of the successful BA output (\protect\subref{fig:SuccessfulBA}) and the angle discretization errors (\protect\subref{fig:BeamRefinementExample}).}
\end{figure}

Let us consider a BS with a co-located radar receiver as well as $\Na$ antennas and $\Nrf$ RF chains. 
Inspired by the IEEE 802.11ad \cite{Kumari_802_11ad_Radar} whose frame structure is shown in \figurename\ref{fig:Frame}, we assume that the BS performs synchronization and initial beam alignment during the short training field (STF) according to the existing methods (e.g. \cite{Heath-beamforming, Xiaoshen}). The output of successful beam alignment is shown in \figurename\ref{fig:SuccessfulBA}, where directional beams connecting the BS and the users have been established. Unfortunately, due to the finite nature of the beam dictionaries used in most BA methods (see e.g. \cite{Heath-SparsePrecoding, JavidimmWave}), the achievable beamforming gain, both at the BS and at the UE, is limited by angle discretization errors as illustrated in \figurename\ref{fig:BeamRefinementExample}. Then, the BS uses the channel estimation field (CEF) to transmit multiple data streams using OFDM to $K\leq\Nrf$ users using narrow beams pointing towards directions $\hat{\phiv} \eqdef [\hat{\phi}_{1}. \dots, \hat{\phi}_{K}]$, where $\hat{\phi}_{k}$ is the angle estimate of user $k$ (constrained to the finite set of angles), while simultaneously processing the backscattered signals at the co-located radar receiver to refine the state estimation.We call this process \textit{beam refinement}. If the beam refinement succeeds, the BS and the users can communicate at higher data rates during the subsequent communication data blocks (BLK). 

\subsection{Channel Model}
We model the time-varying channel response for the $k$-th UE as a line of sight (LOS) channel given by
\begin{align}\label{eq:channel_user_p}
	\Hm^{\rm UE}_{k} (t, \tau) =  h_{k} \av(\theta_{k}) \av^\H(\phi_{k})\delta(\tau-\tau_{k}/2)  e^{j\pi \nu_{k} t}\,,
\end{align}
for $k=1,\dots,K$, where $h_{k}$ is the complex valued channel coefficient, $\theta_{k}$, $\phi_{k}$ are angle of arrival (AoA), angle of departure (AoD), $\tau_{k}/2$ and $\nu_{k}/2$ are one-way delay and one-way Doppler shift respectively. 
Similarly, we model the two-way channel seen at the radar receiver as a $K$-path time-varying channel response given by 
\begin{align}\label{eq:channel}
	\Hm (t, \tau) =  \sum_{k=1}^{K}h'_{k} \av(\phi_{k}) \av^\H(\phi_{k})\delta(\tau-\tau_{k})  e^{j2\pi \nu_{k} t}\,,
\end{align}
where $h'_{k}$ is the channel coefficient corresponding to the two-way channel between the BS and the $k$-th UE. By letting all the antenna architectures be uniform linear arrays (ULA) with half-wavelength inter-element spacing, the array response vectors are given by $[\av(\xi)]_{i} = e^{j\pi(i-1)\sin(\xi)}$, with dimensionality equal to the number of antennas of each device. With some abuse of notation, we used the same name for the array response vector of every device, but the dimensionality should be clear from the context.

\subsection{Signal Model}
\newcommand{\rect}{{\sf rect}}
We consider (OFDM) since it is one the standardized waveforms for mmWave systems and due to its robustness and its ability to deal with time-invariant frequency selective channels. 
In OFDM systems, the total bandwidth $B$ is divided into $M$ subcarriers, i.e., $B=M \Delta f$, where $\Delta f$ [\si{\hertz}] denotes the subcarrier bandwidth. 
For a given maximum Doppler shift $\nu_{\max}$, the subcarrier bandwidth is typically chosen to satisfy\footnote{Note that this approximation can be justified in a number of 
scenarios. For example, consider a scenario inspired by IEEE 802.11p with $f_c$ = \SI{5.89}{\giga\hertz} and the subcarrier spacing $\Delta f$ = \SI{156.25}{\kilo\hertz}. 
This yields $v_{\max} \ll$ \SI{14325}{\kilo\meter/\hour}, which is reasonable even for a relative speed of \SI{400}{\kilo\meter/\hour}. The same holds for IEEE 802.11ad with $f_c$ = \SI{60}{\giga\hertz} and $\Delta f$ = \SI{5.15625}{\mega\hertz} \cite{cordeiro2010ieee}.}
\begin{equation}\label{eq:max-Doppler}
	\nu_{\max}  \ll\Delta f.
\end{equation}

For each OFDM data symbol of duration $T=1/\Delta f $, a cyclic prefix (CP) is appended in order to avoid inter-block interference between the adjacent OFDM symbols, resulting in a total symbol duration of $T_{\rm 0}= T_{\rm cp} + T$.
By considering $N$ symbols, the OFDM frame duration is $T_{f}^{\rm ofdm}=NT_{\rm 0}$. Furthermore, we let the BS apply a transmit beamforming matrix $\Fm(\hat{\phiv}) = [\fv(\hat{\phi}_{1}),\dots,\fv(\hat{\phi}_{K})]\in\CC^{\Na\times K}$ pointing towards the direction of $K$ users in order to obtain beamforming gain. Finally, a reduction matrix $\Um \in \CC^{\Na\times\Nrf}$  and a combining vector $\vv_{k}$ are used respectively at the BS and $k$-th user receiver.

Under this setup, the continuous-time OFDM transmitted signal with CP is given by 
\begin{align}
	\sv(t) = \Fm(\hat{\phiv})\sum_{n=0}^{N-1}\sum_{m=0}^{M-1}  \xv[n, m]\rect_{T_{\rm 0}}(t - n T_{\rm 0})e^{j 2\pi m \Delta f (t-T_{\rm cp}-nT_{\rm o})}\,,
\end{align}
where $\xv[n, m] = [x_{1}[n, m],\dots,x_{K}[n,m]]^\T$ are the information bearing symbols satisfying the average power constraint $\EE[\xv[n,m]\xv^\H[n,m]] = \frac{P_{\rm t}}{K}\Id_{K}$, $\Id_{K}$ is the identity matrix of rank $K$ and $\rect_{T_{\rm 0}}(t)$ is one for $t\in [0, T_{0}]$ and zero elsewhere.

The received backscattered signal in the absence of noise is given by
\begin{align}
	\yv(t) &= \Um^\H\sum_{k=1}^{K}h'_{k}\av(\phi_{k})\av^\H(\phi_{k})\sv(t-\tau_{k})e^{j2\pi\nu_{k}t} \\
	&= \Um^\H\sum_{k=1}^{K}h'_{k}\av(\phi_{k})\av^\H(\phi_{k})\sum_{k'=1}^{K}\fv(\hat{\phi}_{k'})\sum_{n=0}^{N-1}\sum_{m=0}^{M-1}  \xv[n, m]\rect_{T_{\rm 0}}(t - \tau_{k} - n T_{\rm 0}) e^{j2\pi m\Delta f(t - \tau_{k} - T_{\rm cp} - nT_{\rm o}}e^{j2\pi\nu_{k}t}\\
	&\approx \Um^\H\sum_{k=1}^{K}h'_{k}\av(\phi_{k})\av^\H(\phi_{k})\fv(\hat{\phi}_{k})\sum_{n=0}^{N-1}\sum_{m=0}^{M-1}  x_{k}[n, m]\rect_{T_{\rm 0}}(t - \tau_{k} - n T_{\rm 0}) e^{j2\pi m\Delta f(t - \tau - T_{\rm cp} - nT_{\rm o}}e^{j2\pi\nu_{k}t},
\end{align}
where the last step follows from the approximation  $|\av^\H(\phi_{k})\fv(\hat{\phi}_{k'})| \approx 0$ for $k' \neq k$, which is accurate in massive MIMO systems when the users are spatially separated \cite{mMIMO-fundamentals}.

After standard OFDM processing (see e.g. \cite{sturm2011waveform}) and including noise, the samped signal is given by
\begin{align}
	\yv[n, m] &= \Um^\H\left(\sum_{k=1}^{K}h'_{k}e^{j2\pi(nT_{\rm 0}\nu_{k} - m\Delta f\tau_{k})}\av(\phi_{k})\av^\H(\phi_{k})\fv(\hat{\phi}_{k})x_{k}[n, m] + \wv[n, m]\right)\\
	&= \Um^\H\left(\sum_{k=1}^{K}h'_{k}g_{t, k}\av(\phi_{k})\tilde{x}_{k}[n, m] + \wv[n, m]\right),
\end{align}
where $\wv[n, m] \in \CC^{\Na}$ is assumed to be spatially and temporally white Gaussian noise with variance $\sigma_{n}^{2}$ and we have defined  $g_{{\rm t}, k} \eqdef \av^\H(\phi_{k})\fv(\hat{\phi}_{k})$ and $\tilde{x}_{k}[n, m] \eqdef x_{k}[n, m]e^{j2\pi(nT_{\rm 0}\nu_{k} - m\Delta f\tau_{k})}$. Note that $g_{{\rm t}, k}$ depends on the accuracy of the angle estimates $\hat{\phi}_{k}$, but we do not make the dependency explicit in \eqref{eq:rx_sig_user} for the sake of simplicity. 

Finally, it will be shown in the next section that, in order to refine $\hat{\phi}_{k}$ for some specific user index $k$, the matrix $\Um$ is designed such that $\|\Um^\H\av(\phi_{k'})\|_{2} \approx 0$ for $k'\neq k$. Under this assumptions, the channel output at the BS when refining $\hat{\phi}_{k}$ is given by
\begin{align}\label{eq:RadarRxUser}
	\yv[n, m]  \approx\Um^\H(h'_{k}g_{{\rm t}, k}\av(\phi_{k})\tilde{x}_{k}[n, m] + \wv[n, m]),
\end{align}

Under similar assumptions, the received signal at the $k$-th UE after FFT processing is given by
\begin{align}\label{eq:rx_sig_user}
	y^{\rm UE}_{k}[n, m] \approx h_{k}g_{{\rm t}, k}g_{{\rm r}, k}x_{k}[n,m] e^{j2\pi \left(n T_{\rm 0} \frac{\nu_{k}}{2} - {m} \Delta f \frac{\tau_{k}}{2}\right)}+w[n, m],
\end{align}
where we further introduced $ g_{{\rm r}, k} \eqdef \vv^\H_{k}\av(\theta_{k})$, and $w[n, m]$ is white Gaussian noise of the same power as in the BS receiver. 

\section{Beam Refinement}\label{ssec:DOA}

\begin{figure}[!t]
	\centering
	\includegraphics[width=.5\textwidth]{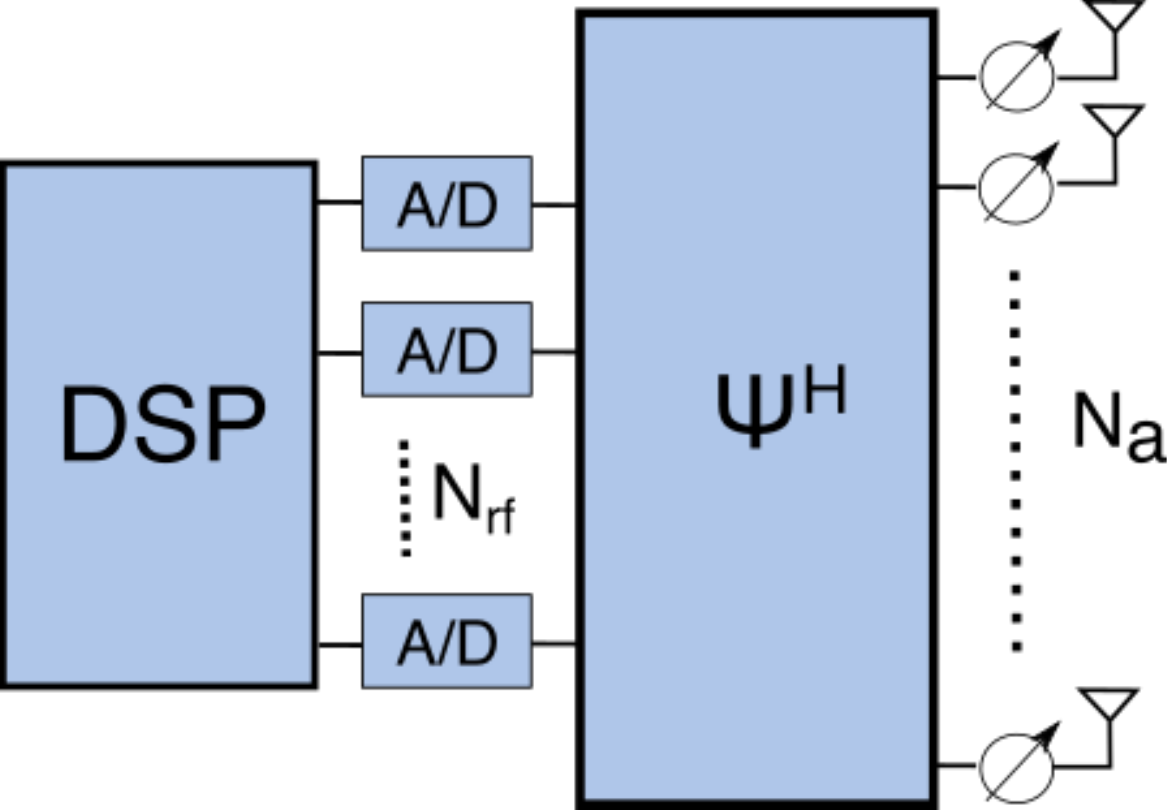}
	\caption{Block diagram of the receiver architecture including a network of tunable phase shifters and a fixed set of spatial filters $\Psim\in\CC^{\Na\times\Nrf}$.}
	\label{fig:receiver}
\end{figure}

From now on we focus on the beam refinement of a generic UE based on the observation \eqref{eq:RadarRxUser}, and will therefore stop using the subindex $k$ to reduce notation clutter. Notice that if multiple angles are to be refined, our approach can be only be applied sequentially. 

In order to use the coarse angle information $\hat{\phi}$ in a hardware efficient way, we decompose $\Um$ as 
\begin{align}
	\Um(\hat{\phi}) = \Dm(\hat{\phi})\Psim,
\end{align}
where $\Psim\in\CC^{\Na\times\Nrf}$ is a fixed set of orthogonal beamforming vectors pointing towards the broadside direction (i.e. \ang{0} or low pass spatial filters) and $\Dm(\hat{\phi})$ is a diagonal matrix where the nonzero entries have unit magnitude, representing a network of phase shifters tuned according to $\hat{\phi}$. This architecture is shown in \figurename\ref{fig:receiver}. In particular, we let $[\Dm(\hat{\phi})]_{i,i} = e^{-j(i-1)\pi\sin(\hat{\phi})}$, such that $\Dm(\hat{\phi})$ effectively \textit{demodulates} angles around $\hat{\phi}$ into broadside. Indeed, we have  
\begin{align}\label{eq:306}
	\left[\Dm^\H(\hat{\phi})\av(\phi)\right]_{i} = e^{-j(i-1)\pi(\sin(\phi)-\sin(\hat{\phi}))},\;\; i=1,\dots,\Na
\end{align}
which shows that $\Dm^\H(\hat{\phi})\av(\phi) = \av\left(\sin^{-1}(\sin(\phi) - \sin(\hat{\phi}))\right)$. If we consider $\phi=\hat{\phi} + \epsilon$ for some small $\epsilon$ (expressed in radians), 
\begin{align}
	\sin(\phi) - \sin(\hat{\phi}) &= \sin(\hat{\phi} + \epsilon) - \sin(\hat{\phi}) \nonumber \\
	&\stackrel{(a)}=\sin(\hat{\phi})\cos(\epsilon) + \cos(\hat{\phi})\sin({\epsilon}) - \sin(\hat{\phi})  \nonumber \\
	&\stackrel{(b)}\approx \sin(\hat{\phi})\left(1-\frac{\epsilon^{2}}{2}\right) + \epsilon\cos(\hat{\phi}) - \sin(\hat{\phi})\label{eq:Taylor_approx} \nonumber \\
	&=\epsilon\cos(\hat{\phi}) + \Oc(\epsilon^{2}),
\end{align}
where (a) follows from the trigonometric identity; (b) follows from the Taylor expansion of functions $\sin, \cos$. Since $|\epsilon\cos(\hat{\phi})| \leq \epsilon \ll 1$ holds, it readily follows 
\begin{align}
	\left|\sin^{-1}(\epsilon\cos(\hat{\phi})) \right| \leq \left|\sin^{-1}(\epsilon)\right| \approx |\epsilon|.
\end{align}
Therefore, we have from \eqref{eq:306} that $\Dm^\H(\hat{\phi})\av(\phi)=\av(\epsilon')$ with $|\epsilon'|\leq|\phi-\hat{\phi}|$, so that signals coming from directions around $\hat{\phi}$ are seen as signals coming from broadside after the phase shifter network $\Dm(\hat{\phi})$.

\begin{figure}[!h]
	\centering
	\begin{subfigure}{.3\textwidth}
		\includegraphics[width=\textwidth]{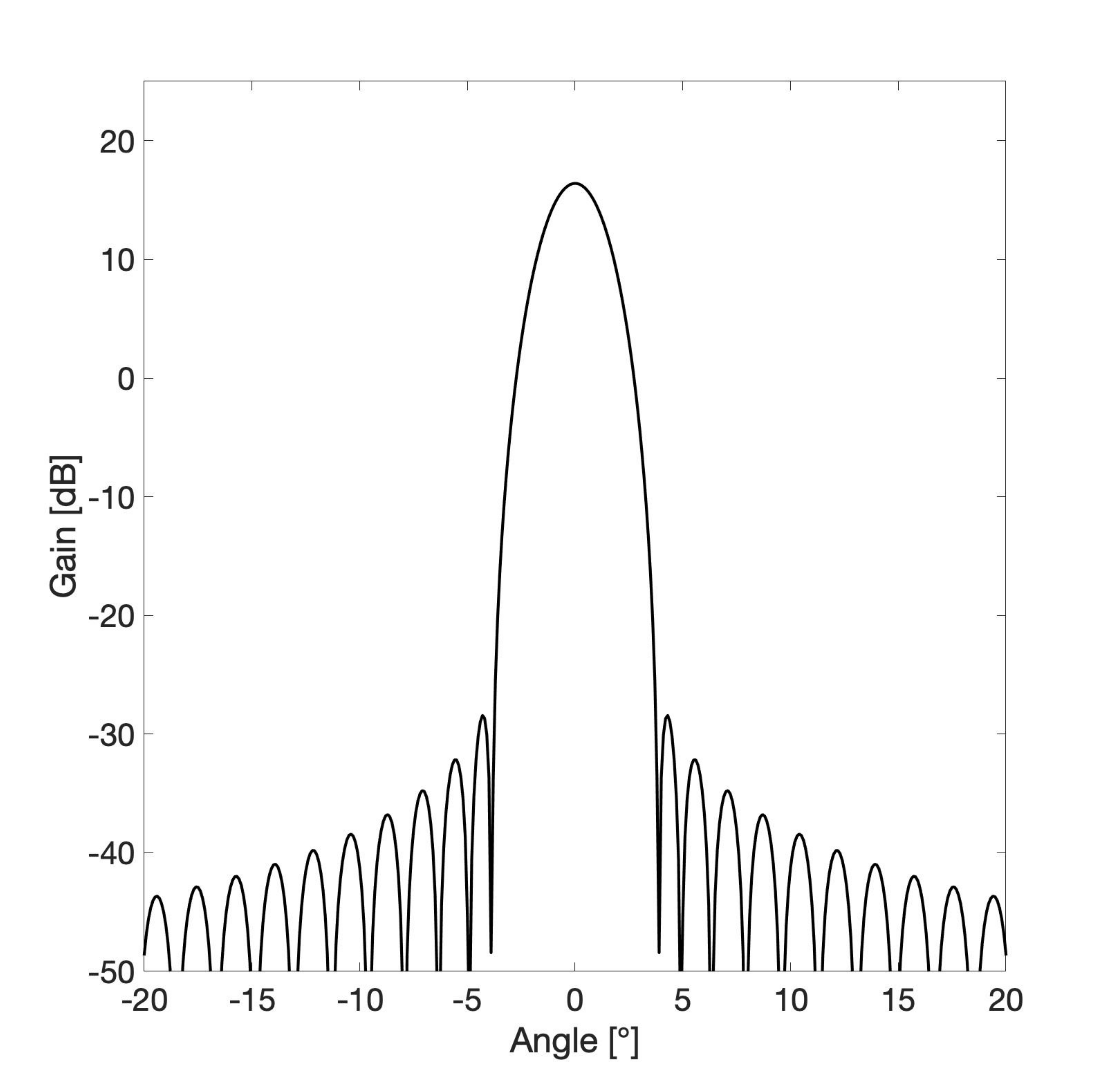}
		\caption{First Slepian sequence}
	\end{subfigure}
	\begin{subfigure}{.3\textwidth}
		\includegraphics[width=\textwidth]{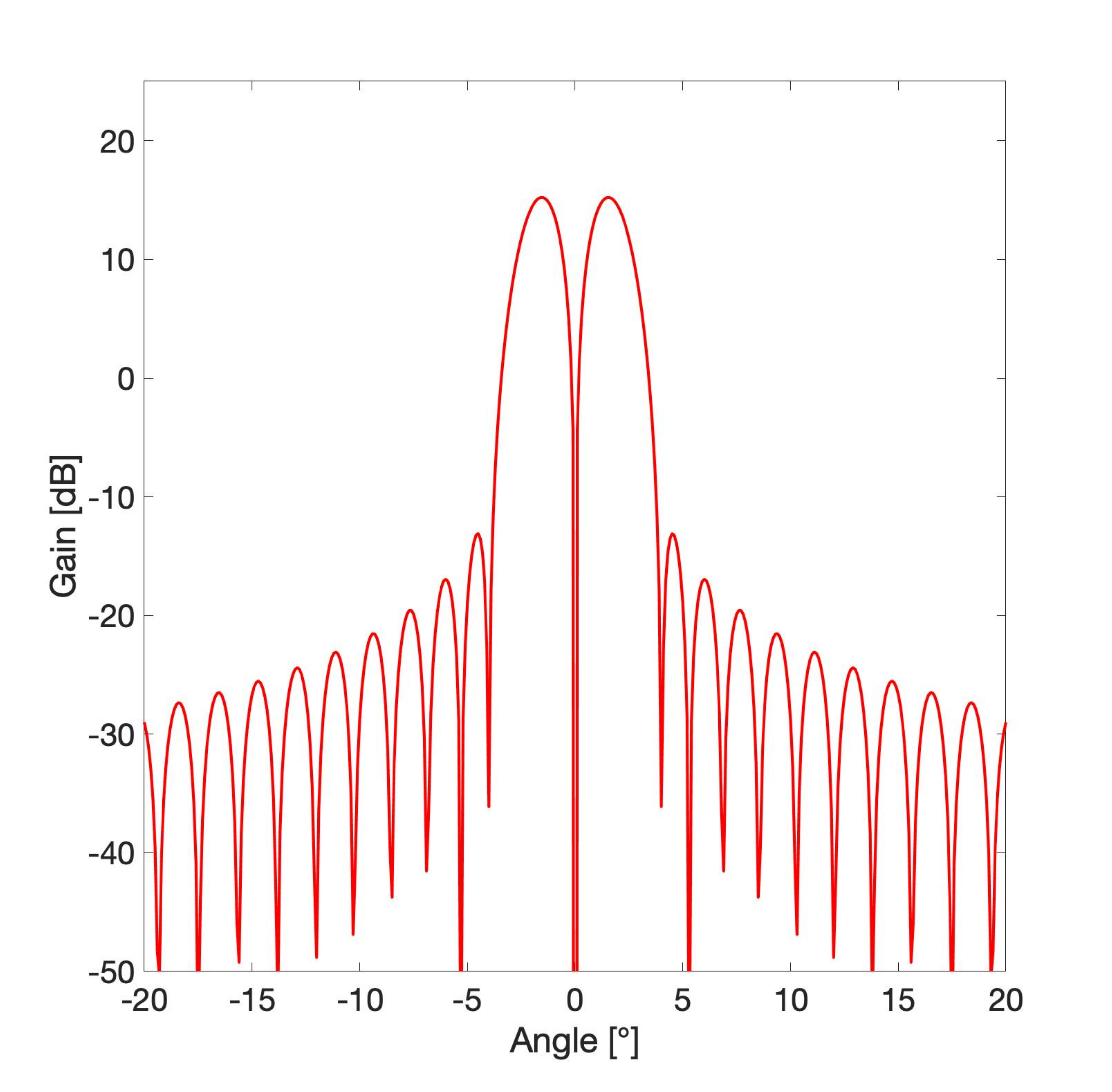}
		\caption{Second Slepian sequence}
	\end{subfigure}
	\begin{subfigure}{.3\textwidth}
		\includegraphics[width=\textwidth]{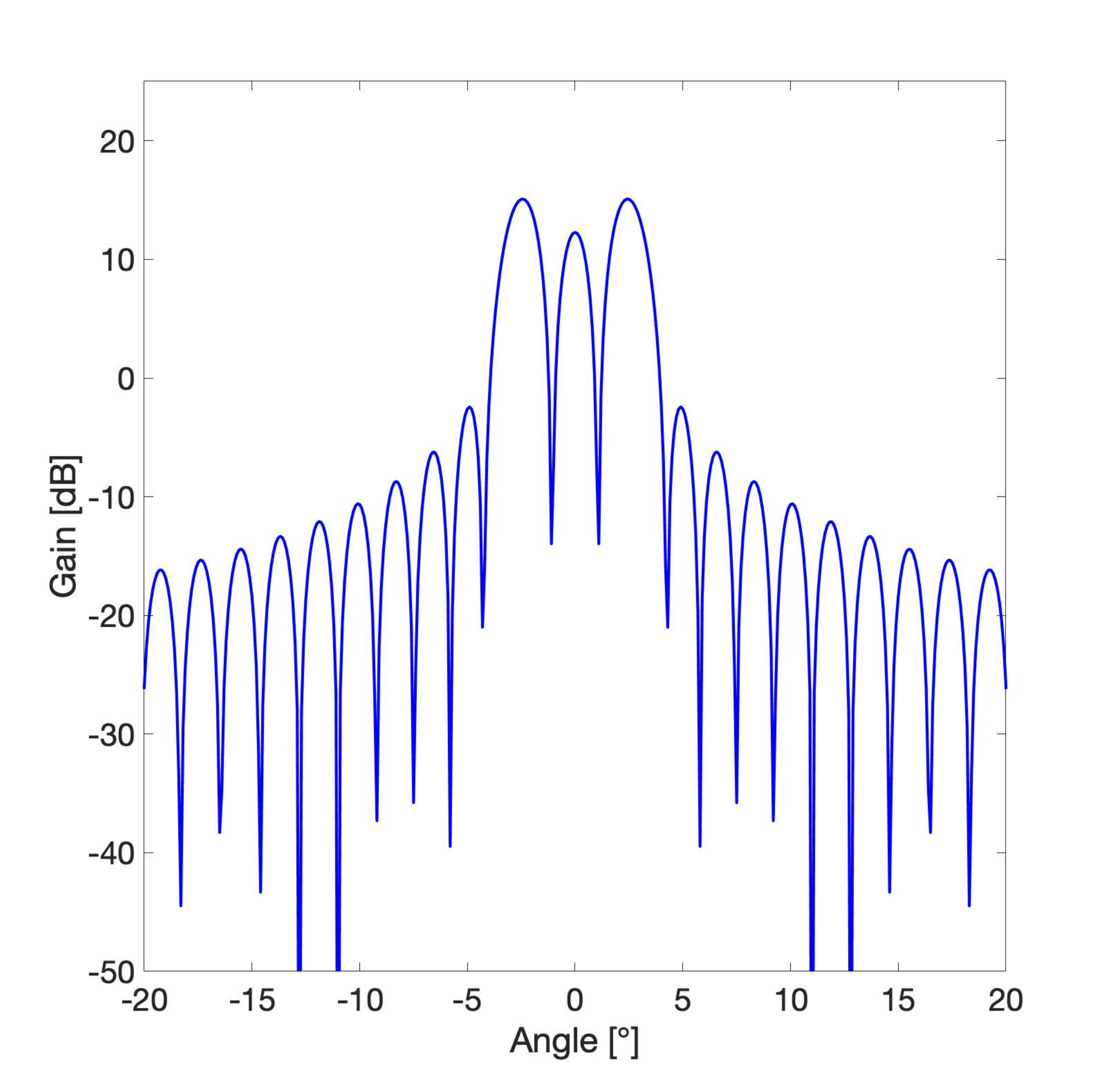}
		\caption{Third Slepian sequence}
	\end{subfigure}	
	\caption{Beam patterns generated by the first three Slepian sequences when $\Na=64$ and $\beta=4/\Na$.}
	\label{fig:SlepianSequences}
\end{figure}

In order to process the broadside incoming signal, we make use of Slepian sequences \cite{StoicaSpectralAnalysis}, which are a set of orthonormal vectors whose spatial spectrum is maximally concentrated around broadside. In particular, the first Slepian sequence $\psiv_{1}$ is given by the solution of
\begin{equation}\label{eq:DPSS_Formulation}
	\begin{aligned}
			\psiv_{1} =  &\underset{\psiv\in\CC^{\Na}}{\arg\max}\;\; & &\frac{1}{2\pi}\int_{-\beta\pi}^{\beta\pi}\left|\psiv^\H\av(\gamma)\right|^{2}d\gamma  \\
		&\text{subject to} & & \frac{1}{2\pi}\int_{-\pi}^{\pi}\left|\psiv^\H\av(\gamma)\right|^{2}d\gamma = 1,
	\end{aligned}
\end{equation}

where $\gamma = \pi\sin(\phi)$ and $\beta \geq 1/\Na$ is a user-defined parameter controlling the beamwidth of the sequence. The solution of problem \eqref{eq:DPSS_Formulation} is known to be given by the eigenvector corresponding to the largest eigenvalue of $\Gammam$, where\cite{StoicaSpectralAnalysis}
\begin{align}
\Gammam = \frac{1}{2\pi}\int_{-\beta\pi}^{\beta\pi}\av(\gamma)\av^\H(\gamma)d\gamma.
\end{align}
Furthermore, the eigenvector $\psiv_{i}$ corresponding to the $i$-th largest eigenvalue of $\Gammam$ is called the $i$-th Slepian sequence. If $\Nrf \leq \beta\Na$, the first $\Nrf$ Slepian sequences yield a set of mutually orthogonal beamforming vectors that reject angles bigger than $\sin^{-1}(\beta)\approx\beta$, and therefore we construct $\Psim = [\psiv_{1},\dots, \psiv_{\Nrf}]$. Note that this justifies our assumption that the radar can focus on the return of a single user. For the sake of illustration, the angular response of the 3 first Slepian sequences for $\Na=64$ and $\beta=4/\Na$ is shown in \figurename\ref{fig:SlepianSequences}. 

Assuming that the signal $x[n,m]$ is uncorrelated with the noise, the sample covariance matrix of the received signal $\yv[n, m]$ is given by
\begin{align}\label{eq:BS_MUSIC}
	\hat{\Rm} &= \frac{1}{NM}\sum_{n=0}^{N-1}\sum_{m=0}^{M-1}\yv[n, m]\yv^\H[n, m]  \nonumber\\
	&\approx |h'|^{2}|g_{\rm t}|^{2}\Psim^{\H}\av(\phi')\av^\H(\phi')\Psim\frac{1}{NM}\sum_{n,m}|\tilde{x}[n,m]|^2 
	 + \Psim^\H\Dm^\H(\hat{\phi})\left(\frac{1}{NM}\sum_{n,m}\wv[n, m]\wv^\H[n, m]\right)\Dm(\hat{\phi})\Psim \nonumber\\
	&\approx \frac{P_{\rm t}}{K}|h'|^{2}|g_{\rm t}|^{2}\Psim^\H\av(\phi')\av^\H(\phi')\Psim + \sigma_{n}^{2}\Id_{\Nrf},
\end{align}
where $\phi' \eqdef \sin^{-1}(\sin(\phi) -\sin(\hat{\phi}))$. The angle $\phi'$ can be readily obtained from \eqref{eq:BS_MUSIC} by applying beamspace MUSIC \cite{BeamspaceMUSIC} to $\hat{\Rm}$. Finally, the refined angle estimate $\check{\phi}$ is computed as $\check{\phi} = \sin^{-1}(\sin(\hat{\phi}) + \sin(\phi'))$.

We can use the refined angle estimate $\check{\phi}$ in order to obtain the estimates of range and velocity useful for beam tracking algorithms. To do that, we formulate the delay and Doppler estimation problem with the non linear least squares minimization
\begin{align}\label{eq:NLS}
	(\hat{\tau}, \hat{\nu}) = \underset{(\tau, \nu)}{\arg\min}\;\; \sum_{n,m}\left\|\yv - h'g_{\rm t}\Um^\H\av(\check{\phi})\tilde{x}[n, m]\right\|_{2}^{2}\;\;,
\end{align}
where the dependency with the parameters $\tau$ and $\nu$ is only present in the term $\tilde{x}[n, m]$. By doing some algebra, we can rewrite expression \eqref{eq:NLS} as
\begin{align}\label{eq:NLS2}
	&(\hat{\tau}, \hat{\nu}) = \underset{(\tau, \nu)}{\arg\min}\;\; \sum_{n,m}\yv^\H[n, m]\left(\Id_{\Nrf} + \Bm\right)\yv[n, m] 
	+\left|h'g_{\rm t}\tilde{x}[n, m] - \frac{\av^\H(\check{\phi})\Um\yv[n, m]}{\av^\H(\check{\phi})\Um\Um^\H\av(\check{\phi})}\right|^{2}\left(\av^\H(\check{\phi})\Um\Um^\H\av(\check{\phi})\right),
\end{align}
where $\Bm = \frac{\Um^\H\av(\check{\phi})\av^\H(\check{\phi})\Um}{\av^\H(\check{\phi})\Um\Um^\H\av(\check{\phi})}$ is a matrix that does not depend on $\tau$ and $\nu$. From \eqref{eq:NLS2}, it is clear that problem \eqref{eq:NLS} can be simplified as
\begin{align}\label{eq:NLS-intermediate}
	(\hat{\tau}, \hat{\nu}) &= \underset{(\tau, \nu)}{\arg\min}\;\; \sum_{n,m}\left|h'g_{\rm t}\tilde{x}[n, m] - \frac{\av^\H(\check{\phi})\Um\yv[n, m]}{\av^\H(\check{\phi})\Um\Um^\H\av(\check{\phi})}\right|  \nonumber\\
	&= \underset{(\tau, \nu)}{\arg\min}\;\; \sum_{n,m}\left|y'[n, m]-h'g_{\rm t}e^{j2\pi(nT_{\rm 0}\nu - m\Delta f\tau)}x[n, m] \right|,
\end{align}
where $y'[n, m] = \frac{\av^\H(\check{\phi})\Um\yv[n, m]}{\av^\H(\check{\phi})\Um\Um^\H\av(\check{\phi})}$. Problem \eqref{eq:NLS-intermediate} is analogous to the maximum likelihood estimator for sinusoid parameters \cite{richards2005fundamentals} which can be efficiently solved by being casted as
\begin{align}\label{eq:NLS-Last}
	&(\hat{\tau}, \hat{\nu}) = \underset{(\tau, \nu)}{\arg\max}\;\; \sum_{n,m}y'[n,m]x^*[n,m]e^{-j2\pi(nT_{\rm 0}\nu - m\Delta f\tau)}.
\end{align}
The objective function in \eqref{eq:NLS-Last} can be computed just by means of FFTs, which enables fast estimation.

Finally, we obtain the range and velocity estimates from 
\begin{align}
	\hat{d} = \frac{c\hat{\tau}}{2} \qquad \hat{v} = \frac{\hat{\nu}\lambda}{2},
\end{align}
where $\lambda$ denotes the wavelength and $c$ is the speed of light. 
\section{Numerical Results}

We set the number of antennas to $\Na=64$ and the number of RF chains to $\Nrf=4$. Based on the IEEE 802.11ad standard, the carrier frequency is chosen to be $f_{c} = $ \SI{60}{\giga\hertz}. For a line of sight model, the SNR at the communications receiver is given by
\begin{align}
	\SNR_{\rm UE} = \frac{|h|^{2}|g_{\rm t}g_{\rm r}|^{2}P_{\rm t}}{K\sigma_{n}^{2}},
\end{align}
where  $|h|^{2} = \lambda^{2}/(4\pi d)^{2}$ is the attenuation for a given distance $d$ in [m] from the BS, $g_{\rm t}$ assumes pointing towards $\hat{\phi}$, and is therefore dependent on the angle discretization error. 
The SNR at each receive antenna of the BS (before the reduction matrix $\Um$) is given by
\begin{align}\label{eq:SNR_BS}
	 \SNR_{\rm BS} = \frac{|h'|^{2}|g_{\rm t}|^{2}P_{\rm t}}{\Nrf\sigma_{n}^{2}} = \SNR_{\rm UE}\frac{\sigma_{\rm rcs}}{4\pi d^{2}|g_{\rm r}|^{2}},
\end{align}
where we let $|h'|^{2} = \frac{\lambda^{2}\sigma_{{\rm rcs}, k}}{(4\pi)^{3}d^{4}}$ for a given radar cross section (RCS) of $\sigma_{{\rm rcs}}$ \cite{richards2005fundamentals}, and assumed for simplicity that the noise power is the same as in the user. The parameters inspired by IEEE802.11ad are summarized in Table \ref{tab:params}. Note that we do not consider the antenna gain at the BS receiver in \eqref{eq:SNR_BS} in order to compare the dependency of the estimation performance with the angle discretization error.

\begin{table}
	\centering
	\caption{System parameters}
	\label{tab:params}
	\begin{tabular}{|c|c|}
		\hline
		N = 16 & M = 512 \\
		\hline
		$\Delta f$ = \SI{1}{\mega\hertz} & $f_{c} = $ \SI{60}{\giga\hertz} \\
		\hline
		$\Na = $ 64 & $\Nrf = $ 4 \\
		\hline
		$\sigma_{\rm rcs} = $ \SI{20}{\dBsm} & $|g_{\rm r}|^{2} = $ 4 \\
		\hline
		$d = $ \SI{40}{\meter} & $\beta = \Nrf/\Na$\\
		\hline
	\end{tabular}
\end{table}

\begin{figure}[!t]
	\centering
	\includegraphics[width=.75\textwidth]{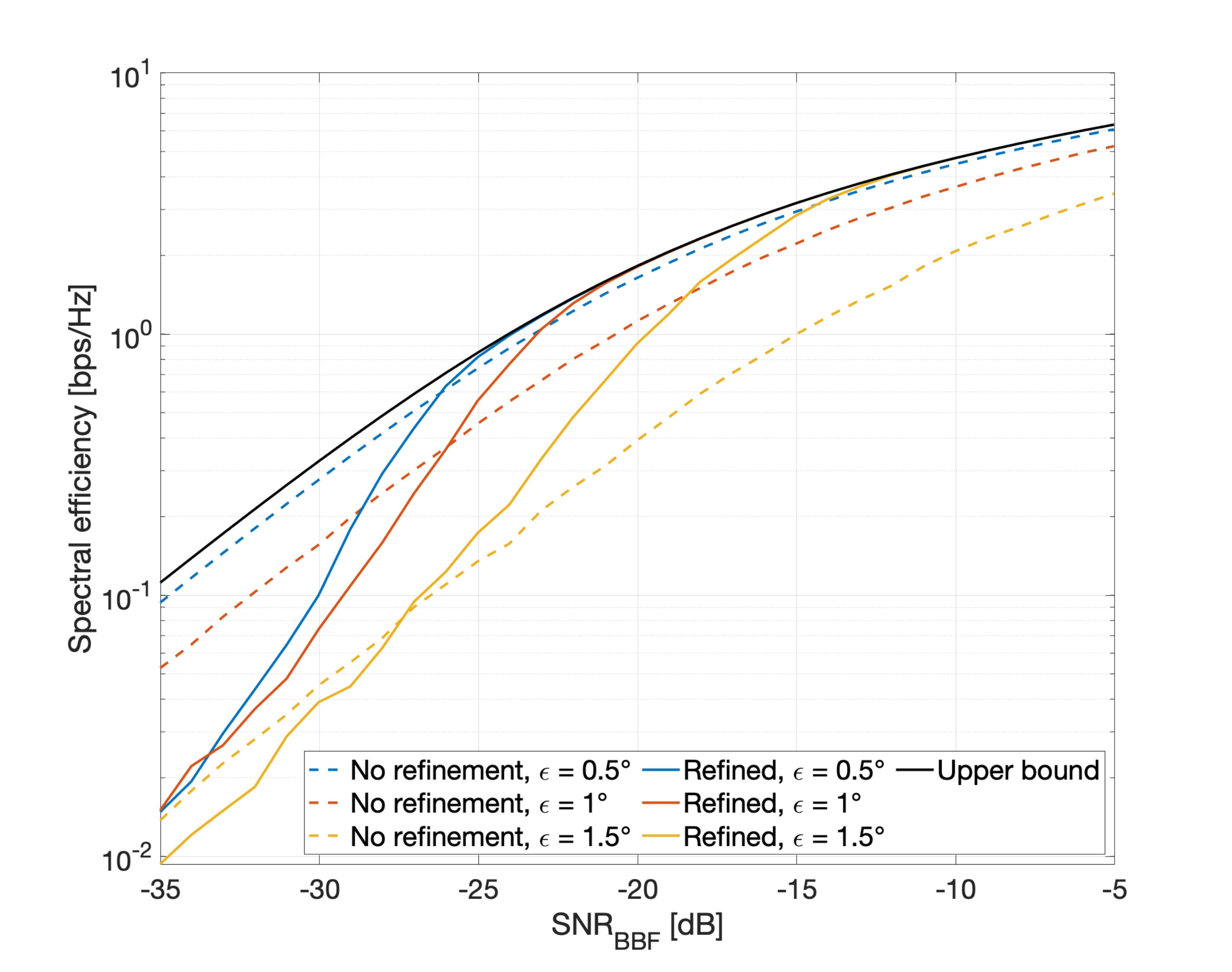}
	\caption{Spectral efficiency as a function of the SNR before beamforming at the communications receiver with and without beam refinement for different angle discretization errors $\epsilon$.}
	\label{fig:spectral_eff}
\end{figure}

In order to quantify the gain of the beam refinement, we evaluate the achievable spectral efficiency, computed as $\log_{2}\left(1 + \SNR_{\rm UE}\right)$, as a function of the SNR at the user receiver before beamforming, that is,
\begin{align}
	\SNR_{\rm BBF} = \frac{|h|^{2}P_{\rm t}}{K\sigma_{n}^{2}}.
\end{align}
This result is shown in \figurename~\ref{fig:spectral_eff} for different angle discretization errors $\epsilon$ along with the `no refinement' case corresponding to the coarse angle estimation. Clearly, for small angle discretization errors (e.g., \ang{0.5}), the gain in the direction of the user is close to the maximum and therefore the improvement in spectral efficiency when using the proposed beam refinement approach is not significant. However, for larger angle discretization errors (e.g. \ang{1.5}), we can obtain improvements of several \si{bps/\hertz}, especially in the high SNR range. We can also notice in the figure how in very noisy conditions the refinement fails and the non refined beam achieves better performance. The distribution of the discretization error $\epsilon$ depends on the type of beam alignment performed before beam refinement. However, notice that the separation between the two closest beams of a DFT codebook for a 64 antenna array is approximately \ang{2}, so discretization errors in the order of \ang{1} are reasonable. 

\begin{figure}
	\centering
	\begin{subfigure}{.45\textwidth}
		\includegraphics[width=\textwidth]{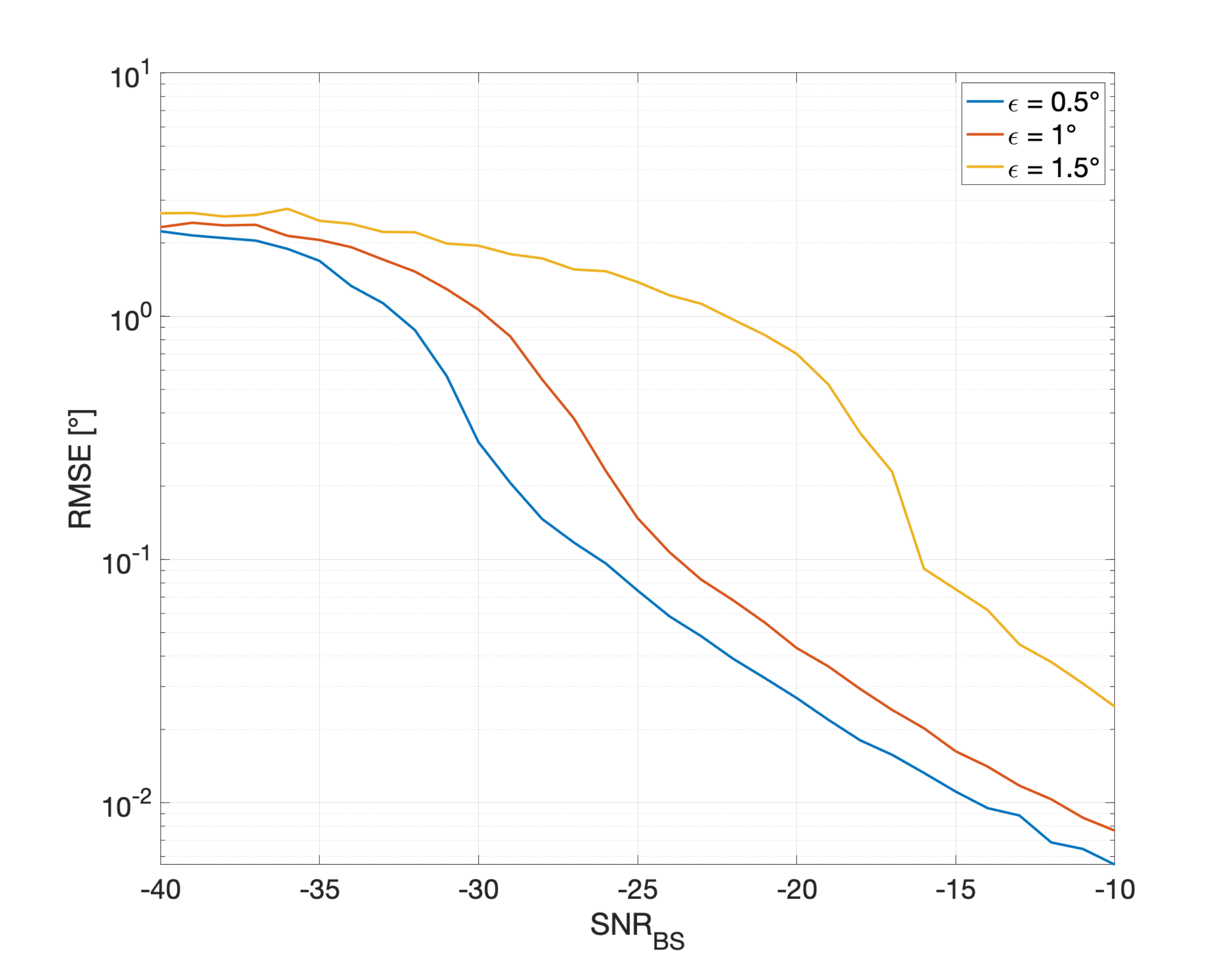}
		\caption{RMSE of the angle estimate}
		\label{fig:RMSE_angle}
	\end{subfigure}
	\begin{subfigure}{.45\textwidth}
		\includegraphics[width=\textwidth]{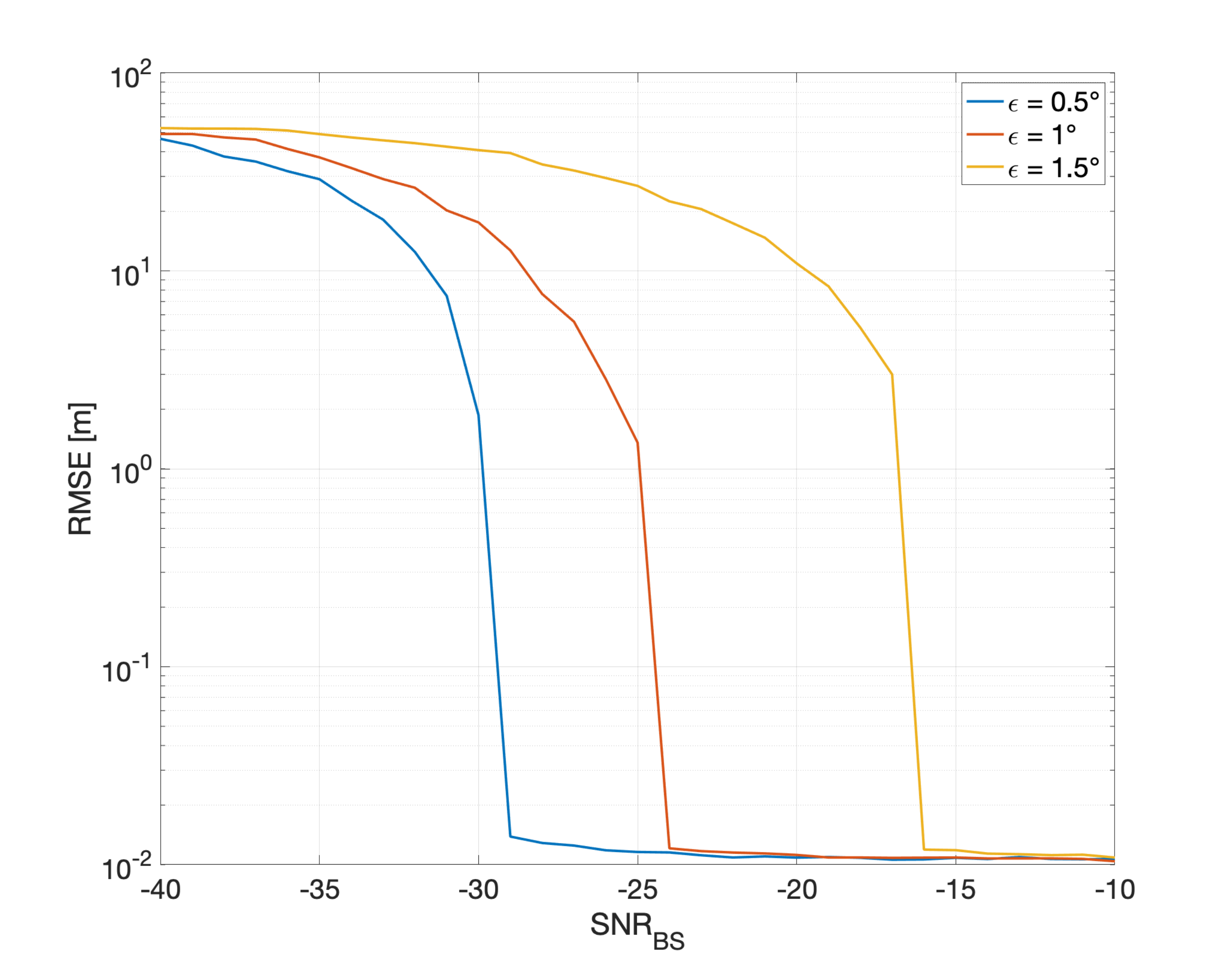}
		\caption{RMSE of the range estimate}
	\end{subfigure}
	\begin{subfigure}{.5\textwidth}
		\includegraphics[width=\textwidth]{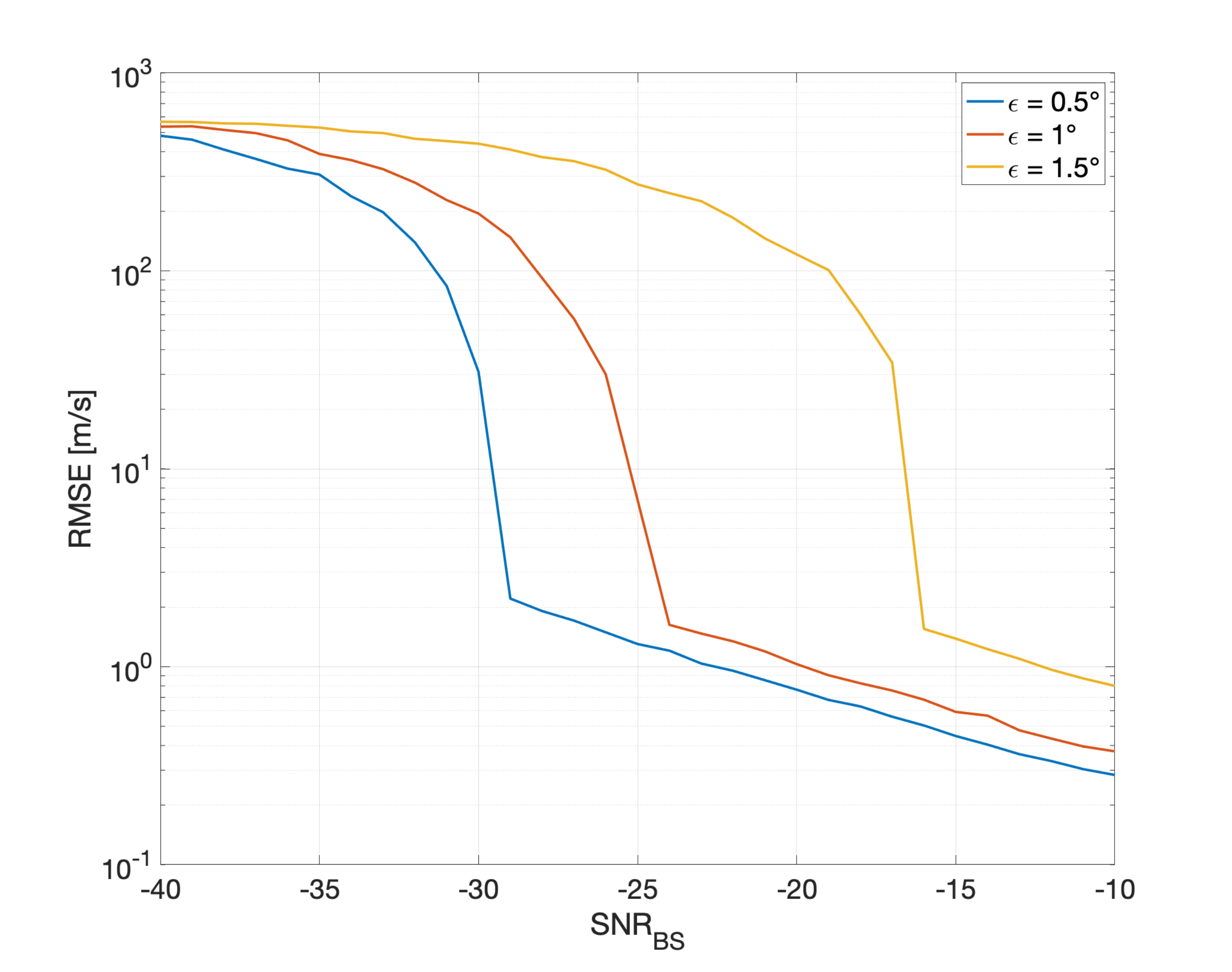}
		\caption{RMSE of the velocity estimate}
	\end{subfigure}
	\caption{RMSE of the different estimators for a given user for different values of the angle discretization error.}
	\label{fig:RMSE}
\end{figure}

Next, we look at the root mean square error (RMSE) of the angle, range and velocity estimates, as shown in \figurename~\ref{fig:RMSE}. The angle RMSE in \figurename~\ref{fig:RMSE_angle} shows that we are able to obtain angle estimation accuracy far beyond the one resulting from using discretized grids. In addition, the range and velocity estimates show reasonable performance to be used as the initial state in beam tracking algorithms.

\section{Conclusions}

In this work, we presented an OFDM-based beam refinement and state acquisition method to exploit opportunistically the presence of existing initial beam alignment protocols. By using a simple hybrid analog architecture to adapt the phase shifters to available coarse angle estimates, the proposed method is able to provide accurate estimation of AoA, hence improve significantly the spectral efficiency. The refined estimate of other parameters will be useful for beam tracking applications in a dynamic environment with user mobility. However, this topic is out of the scope of the present paper and it is left for future investigation. 


\begin{thebibliography}{10}
\providecommand{\url}[1]{#1}
\csname url@samestyle\endcsname
\providecommand{\newblock}{\relax}
\providecommand{\bibinfo}[2]{#2}
\providecommand{\BIBentrySTDinterwordspacing}{\spaceskip=0pt\relax}
\providecommand{\BIBentryALTinterwordstretchfactor}{4}
\providecommand{\BIBentryALTinterwordspacing}{\spaceskip=\fontdimen2\font plus
\BIBentryALTinterwordstretchfactor\fontdimen3\font minus
  \fontdimen4\font\relax}
\providecommand{\BIBforeignlanguage}[2]{{%
\expandafter\ifx\csname l@#1\endcsname\relax
\typeout{** WARNING: IEEEtran.bst: No hyphenation pattern has been}%
\typeout{** loaded for the language `#1'. Using the pattern for}%
\typeout{** the default language instead.}%
\else
\language=\csname l@#1\endcsname
\fi
#2}}
\providecommand{\BIBdecl}{\relax}
\BIBdecl

\bibitem{Heath-SparsePrecoding}
O.~E. {Ayach}, S.~{Rajagopal}, S.~{Abu-Surra}, Z.~{Pi}, and R.~W. {Heath},
  ``Spatially sparse precoding in millimeter wave {MIMO} systems,''
  \emph{{IEEE} Trans. Wireless Commun.}, vol.~13, no.~3, pp. 1499--1513, 2014.

\bibitem{Xiaoshen}
X.~{Song}, S.~{Haghighatshoar}, and G.~{Caire}, ``Efficient beam alignment for
  millimeter wave single-carrier systems with hybrid {MIMO} transceivers,''
  \emph{{IEEE} Trans. Wireless Commun.}, vol.~18, no.~3, pp. 1518--1533, 2019.

\bibitem{Heath-beamforming}
A.~{Alkhateeb}, O.~{El Ayach}, G.~{Leus}, and R.~W. {Heath}, ``Channel
  estimation and hybrid precoding for millimeter wave cellular systems,''
  \emph{IEEE J Sel Top Signal Process}, vol.~8, no.~5, pp. 831--846, 2014.

\bibitem{BeamTracking1}
F.~Liu, W.~Yuan, C.~Masouros, and J.~Yuan, ``Radar-assisted predictive
  beamforming for vehicular links: Communication served by sensing,''
  \emph{{IEEE} Trans. Wireless Commun.}, vol.~19, no.~11, pp. 7704--7719, 2020.

\bibitem{JRC-survey}
F.~Liu, C.~Masouros, A.~P. Petropulu, H.~Griffiths, and L.~Hanzo, ``Joint radar
  and communication design: Applications, state-of-the-art, and the road
  ahead,'' \emph{{IEEE} Trans. Commun.}, vol.~68, no.~6, pp. 3834--3862, 2020.

\bibitem{gonzalez2016radar2}
N.~Gonz{\'a}lez-Prelcic, R.~M{\'e}ndez-Rial, and R.~W. Heath, ``Radar aided
  beam alignment in mmwave {V2I} communications supporting antenna diversity,''
  in \emph{Information Theory and Applications Workshop (ITA), 2016}.\hskip 1em
  plus 0.5em minus 0.4em\relax IEEE, 2016, pp. 1--7.

\bibitem{va2017position2}
V.~Va, T.~Shimizu, G.~Bansal, and R.~W. Heath, ``Position-aided millimeter wave
  {V2I} beam alignment: A learning-to-rank approach,'' in \emph{2017 IEEE 28th
  Annual International Symposium on Personal, Indoor, and Mobile Radio
  Communications (PIMRC)}, 2017, pp. 1--5.

\bibitem{Kumari_802_11ad_Radar}
P.~{Kumari}, J.~{Choi}, N.~{González-Prelcic}, and R.~W. {Heath}, ``{IEEE}
  802.11ad-based radar: An approach to joint vehicular communication-radar
  system,'' \emph{{IEEE} Trans. Veh. Technol.}, vol.~67, no.~4, pp. 3012--3027,
  2018.

\bibitem{JavidimmWave}
S.~{Chiu}, N.~{Ronquillo}, and T.~{Javidi}, ``Active learning and csi
  acquisition for mmwave initial alignment,'' \emph{{IEEE} J. Sel. Areas
  Commun.}, vol.~37, no.~11, pp. 2474--2489, 2019.

\bibitem{cordeiro2010ieee}
C.~Cordeiro, D.~Akhmetov, and M.~Park, ``Ieee 802.11 ad: Introduction and
  performance evaluation of the first multi-gbps wifi technology,'' in
  \emph{Proceedings of the 2010 ACM international workshop on mmWave
  communications: from circuits to networks}.\hskip 1em plus 0.5em minus
  0.4em\relax ACM, 2010, pp. 3--8.

\bibitem{mMIMO-fundamentals}
T.~L. Marzetta, E.~G. Larsson, H.~Yang, and H.~Q. Ngo, \emph{Fundamentals of
  Massive MIMO}.\hskip 1em plus 0.5em minus 0.4em\relax Cambridge University
  Press, 2016.

\bibitem{sturm2011waveform}
C.~Sturm and W.~Wiesbeck, ``Waveform design and signal processing aspects for
  fusion of wireless communications and radar sensing,'' \emph{Proc. IEEE},
  vol.~99, no.~7, pp. 1236--1259, 2011.

\bibitem{StoicaSpectralAnalysis}
P.~Stoica and R.~Moses, \emph{Spectral Analysis of Signals}.\hskip 1em plus
  0.5em minus 0.4em\relax Pearson Prentice Hall, 2005.

\bibitem{BeamspaceMUSIC}
D.~Linebarger, R.~DeGroat, E.~Dowling, and P.~Stoica, ``Constrained beamspace
  music,'' in \emph{1993 IEEE International Conference on Acoustics, Speech,
  and Signal Processing}, vol.~4, 1993, pp. 548--551 vol.4.

\bibitem{richards2005fundamentals}
M.~A. Richards, \emph{Fundamentals of radar signal processing}.\hskip 1em plus
  0.5em minus 0.4em\relax Tata McGraw-Hill Education, 2005.

\end{thebibliography}


\end{document}